# Measure of the Heart: Santorio Santorio and the *Pulsilogium*

Richard de Grijs[1] and Daniel Vuillermin[2]

[1] Kavli Institute for Astronomy and Astrophysics, Peking University, 5 Yi He Yuan Lu, Hai Dian District, Beijing 100871, China; grijs@pku.edu.cn
[2] Institute for Medical Humanities, Peking University Health Science Center, 38 Xue Yuan Lu, Hai Dian District, Beijing 100083, China; dvuillermin@bjmu.edu.cn

*Abstract.* In 1626, the Venetian physician Santorio Santorio published the details of his pulsilogium, a stop clock that could accurately measure one's pulse rate. He applied Galileo Galilei's insights that the frequency of a pendulum's oscillation is inversely proportional to the square root of its length. Santorio's inventions emerged at a time when the natural world and our solar system were beginning to be mapped in remarkable detail. Santorio was a true representative of his era, a period in which scientific developments came in rapid succession and measurements to support hypotheses became the norm.

The heart is a musical organ. The irregularity of one's inhalation and exhalation of air defies musicality, while the involuntary rumbling of moving gas in the intestines is embarrassingly analogous to the timbre of the tuba or trombone. Biomedical terminology and poetry are seemingly antithetical, but of the heart they speak a common musical language of beats, pulses, pounds, throbs, and rhythms. As Jack Kerouac writes, "*It's the beat generation, it's beat, it's the beat to keep, it's the beat of the heart*".[1] Since the 19th Century, the heart's performance has commonly been measured aurally—an intimate ear to the chest, percussion, the ear trumpet, the stethoscope—yet in the 17th Century a patient's pulse rate was calculated by *length*.

In 1626, the Venetian physician and nobleman Santorio Santorio (1561–1636) published the details of his *pulsilogium*, a stop clock that could accurately measure the pulse rate—a device which may well have been the first precision instrument in the history of medicine. Santorio was already a well-known physician in the Republic of Venice when, in 1603, he published a series of books, the *Methodi Vitandorum Errorum Omnium qui in Arte Medica Contigunt Libri Quindecim* (*Methods to Avoid All Errors Occurring in Medical Art, 15 Books*), in which he introduced his invention:

> "*In order to commemorate quickly and exactly my knowledge of the pulse of a patient, I have invented the pulsilogium, which makes it possible to measure exactly the beats of the arteries … and to compare them with the beats of earlier days … With the help of the pulsilogium, we can monitor at what day and at which hour the pulse deviated in intensity and frequency from its natural state.*"[2]

In developing his pulse meter, Santorio applied Galileo Galilei's (1564–1642) insights that the frequency of a pendulum's oscillation is inversely proportional to the square root of its length. The *pulsilogium* was composed of a heavy leaden bob and a silk cord. It worked by adjusting the oscillation frequency by changing the pendulum's length until the periodic oscillation was synchronized with the patient's pulse beat.[3] The pulse rate corresponded to either the position of a knot in the cord alongside a horizontal ruler (see Fig. 1, center) or, in Santorio's original design, to the location of the hand on a dial; the pulse rate was therefore referred to in units of length. Extensive experimentation with his new tool allowed Santorio to derive the circadian rhythm (the 24 hour cycle) of the cardiac frequency. A century later, the French physician François Boissier de Sauvages de Lacroix (1706–1767) still used the device to test his patients' heart rates and cardiac performance.

Santorio and his friend Galileo shared an approach to the study of the body and the natural world, one that was mechanical and quantitative: the inner workings of their inventions and observations became metaphors for Nature's mysteries. Galileo became interested in the periodic motion of a swinging chandelier in the Cathedral of Pisa—presumably when he was distracted during Mass—around 1581 or perhaps in 1582, still during his student days. He is said to have timed the *period* of the chandelier's swing—the time for one complete arc from the point of origin through its maximum swing on the opposite side and back again—with his pulse beats. This enabled him to discover the crucial aspect of pendulums that make them viable timepieces, a property now commonly referred to as *isochronism*: a pendulum's period is, at least for small angles of its swing,[4] approximately independent of its *amplitude*, that is, the maximum extent or angle of the swing.

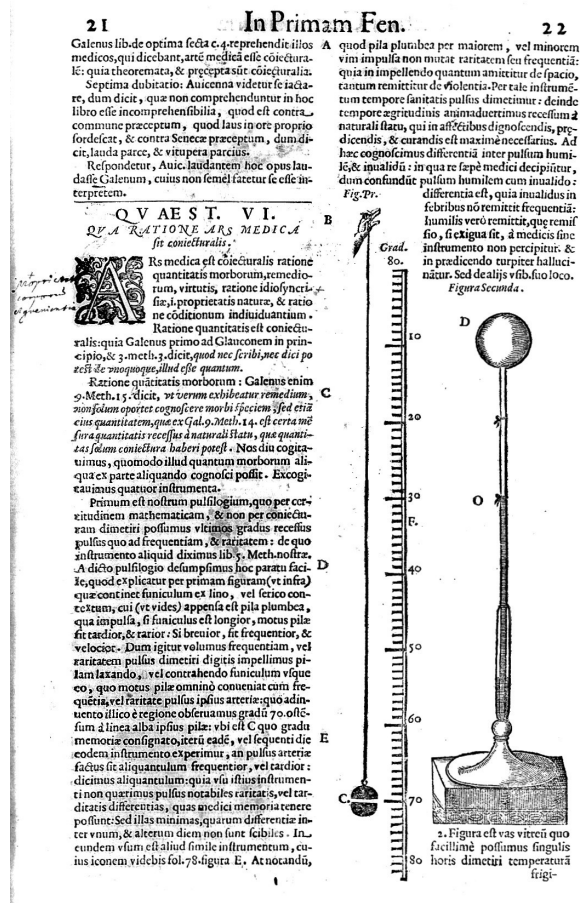

**Figure 1**: *Pulsilogium* (center; line with a weight tied to a finger alongside a ruler) and thermoscope (right). (Sanctorius, S., 1626, *Commentaria in primam Fen primi libri Canonis Avicennae*, Venice: Sarcina, p. 22. Woodcut and text; Credit: *Wellcome Library, London*

Galileo was often preoccupied with thoughts about the practical use of the pendulum. Although he initially employed pendulums to study the laws of motion, he soon proceeded by designing simple timing devices using free-swinging pendulums, including metronomes for students of music. On 29 November 1602, he described the concept of isochronism in a letter to his early patron, the Marquis Guido Ubaldo dal Monte (1545–1607), from Padua. Shortly after Galileo's letter to Dal Monte, Santorio invented the *pulsilogium*.

Although Santorio's medical theories did not deviate significantly from Hippocratic or Galenic practice—based on facilitating a balance of the four bodily fluids or 'humors'—in dealing with his patients' ailments, he relied first and foremost on his sensory experience, both in his teaching at the University of Padua (where he held the Chair of Theoretical Medicine from 1611 until his retirement in 1624) and in relation to his preferred method of investigation. Santorio passionately held on to a view that the fundamental properties of the body were mathematical—in terms of numbers, positions, forms—rather than composed of Aristotelian and Galenic elements and qualities. Like Andreas Vesalius (1514–1564), now often considered the father of modern human anatomy, Santorio was guided by a strong belief in reasoning rather than a reliance on authority.

In his lifetime Santorio was best known for his ground-breaking experimental study of basal metabolism, measuring his bodily weight while at rest, using energy solely to maintain vital cellular activity, respiration, and circulation. Employing a 'balance seat,' a contraption resembling a giant balance, he combined the roles of observer and subject. Over a period of 30 years, Santorio spent much of his time working, eating, and sleeping in his balance chair, weighing his own intake of solid and liquid sustenance, as well as his body's waste products. This led him to eventually conclude that most of his dietary intake was lost from his body through his skin as *perspiratio insensibilis* ('insensible perspiration'):[5] for every eight pounds of food he consumed, his body excreted a mere three pounds. In a translated edition of Santorio's *De Statica Medicina* from 1737, we experience Santorio's interpretation first-hand:

> *"Insensible Perspiration is either made by the Pores of the Body, which is all over perspirable, and cover'd with a Skin like a Net; or it is performed by Respiration through the Mouth, which usually, in the Space of one Day, amounts to about the Quantity of half a Pound, as may plainly be made appear by breathing upon a Glass."*[6]

Although this was not a novel result—it was already known to the ancient Greek physician and surgeon Galen of Pergamon (ca. 129–200/216 CE), arguably the most accomplished medical researcher in the Roman Empire—his experimental work made him the 'father of experimental physiology,' emphasizing the use of mathematics and physics to gain an understanding of physiological processes. He thus represented a 17th Century direction of medical thought known as the *iatrophysical school*.

Santorio's keen sense of mathematics propelled him to invent, either on his own or jointly with Galileo, a number of instruments, including a wind gauge, a hygrometer to measure the ambient humidity, a water-current meter, instruments to extract bladder stones and remove foreign bodies from the ear, a trocar and cannula for surgically draining of fluids from cavities, a device allowing patients to bathe while bed-ridden, and a thermoscope (see Fig. 1, right)—the progenitor of the thermometer in common use today. Descriptions of the thermoscope have been left by both Galileo and Santorio,[7] and the name of the actual inventor has been lost in history, but Santorio is generally credited with applying a numerical scale to the instrument, thus underscoring his quantitative bent once again. The device consisted of an enclosed vessel containing air that contracted or expanded with temperature fluctuations, forcing water to move up or down a tube.

Santorio's inventions emerged at a time when the natural world and our solar system were beginning to be mapped in remarkable detail. The *pulsilogium* is in keeping with Vesalius' maps of the body's internal landscape, which clarified the individual parts of human anatomy with unprecedented precision. The measurement of the body and its internal components would underpin modern physiology, and the *pulsilogium* served as a catalyst for the invention of a vast array of mechanical and digital medical measuring instruments. As an active member of a learned circle in Venice, which boasted Galileo, fra Paolo Sarpi (1552–1623), Hieronymus Fabricius (1537–1619), Giambattista della Porta (1535?–1615), Giovanni Francesco Sagredo (1571–1620), Santorio drew upon this eclectic group of scholars and humanists, where ideas were exchanged freely, leading to some of the most inspirational and revolutionary cross-fertilizations of natural philosophy and medicine. Santorio was a true representative of his era, a period in which scientific developments came in rapid succession and measurements to support hypotheses became the norm.

*Richard de Grijs, Ph.D., is a professor at the Kavli Institute for Astronomy and Astrophysics at Peking University. His forthcoming monograph,* Time And Time Again *(Institute of Physics Publishing, UK), explores the history of the determination of longitude at sea in the $17^{th}$ Century. Daniel Vuillermin, Ph.D., is a lecturer at the Institute for Medical Humanities at the Peking University Health Science Center. He is currently researching a history of medicine and photography in China.*

---

[1] Kerouac, J., 2012, *Desolation Angels*, New York: Penguin Classics.
[2] Sanctorius, S., 1631, *Methodi vitandorum errorum omnium qui in arte medica contingunt*, Geneva: P. Aubertum, p. 289.
[3] Levett, J., Agarwal, G., 1979, The first man/machine interaction in medicine: the *pulsilogium* of Sanctorius, *Med. Instrum.*, 13(1), pp. 61–63.
[4] Galilei, G., 1638, *Discorsi e Dimostrazioni Matematiche Intorno a Due Nuove Scienze*; Galileo claimed isochronicity for all arcs less than 180°, although that claim seems overly optimistic today.
[5] Sanctorius, S., 1614, *De Statica Medicina*, Venice: Niccolò Polo.
[6] Sanctorius, S., 1737, *De Statica Medicina*, London: T. Longman and J. Newton, translation.
[7] Sanctorius, S., 1625, *Commentaria in primam Fen primi libri Canonis Avicennae*, Venice: J. Sarcinam; Santorio also described how the thermoscope ought to be used in studying diseases.

*Figure source*: https://wellcomeimages.org – image L0008488; copyrighted work available under Creative Commons Attribution only licence CC BY 4.0, http://creativecommons.org/licenses/by/4.0/